\documentclass{article}

\usepackage{arxiv}

\usepackage[utf8]{inputenc} 
\usepackage[T1]{fontenc}    
\usepackage{hyperref}       
\usepackage{url}            
\usepackage{booktabs}       
\usepackage{amsfonts}       
\usepackage{nicefrac}       
\usepackage{microtype}      
\usepackage{lipsum}		
\usepackage{graphicx}
\usepackage{natbib}
\usepackage{doi}
\usepackage{amsmath}
\usepackage{indentfirst}
\title{Feasible climate policies in a democracy with a climate-denying party}

\author{ \href{https://orcid.org/0000-0000-0000-0000}{\includegraphics[scale=0.06]{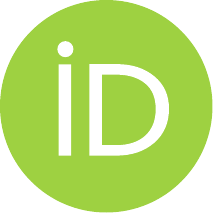}\hspace{1mm}Andrea Di Benedetto}\\
	Institute for Marine and Atmospheric Research (IMAU)\\
        Centre for Complex Systems Studies (CCSS)\\
	Utrecht University\\
	Princetonplein 5, 3584 CC Utrecht, The Netherlands \\
	\texttt{a.dibenedetto@uu.nl} \\
	\And
	\href{https://orcid.org/0000-0000-0000-0000}{\includegraphics[scale=0.06]{orcid.pdf}\hspace{1mm}Claudia E. Wieners} \\
	Institute for Marine and Atmospheric Research (IMAU)\\
        Centre for Complex Systems Studies (CCSS)\\
	Utrecht University\\
	Princetonplein 5, 3584 CC Utrecht, The Netherlands \\
 	\And
	\href{https://orcid.org/0000-0000-0000-0000}{\includegraphics[scale=0.06]{orcid.pdf}\hspace{1mm}Anna S. von der Heydt} \\
	Institute for Marine and Atmospheric Research (IMAU)\\
        Centre for Complex Systems Studies (CCSS)\\
	Utrecht University\\
	Princetonplein 5, 3584 CC Utrecht, The Netherlands \\
}

\renewcommand{\shorttitle}{\textit{arXiv} Template}

\hypersetup{
pdftitle={A template for the arxiv style},
pdfsubject={q-bio.NC, q-bio.QM},
pdfauthor={David S.~Hippocampus, Elias D.~Striatum},
pdfkeywords={First keyword, Second keyword, More},
}

\begin{document}
\maketitle

\begin{abstract}
Climate policy has become increasingly politicized in many countries including the US, with some political parties unwilling to pursue strong measures. Therefore, to be successful in mitigation, climate policies must be politically feasible. Currently, climate mitigation pathways are explored in so-called Integrated Assessment Models (IAMs) which evaluate climate policies from an economic perspective, typically focusing on cost-effectiveness and overlooking transition costs. However, the economy is intertwined with the political system, in which policymakers impose economic policies, but are (in democracies) dependent on public opinion, which in turn can be influenced by economic performance. In cases where some parties are much less ambitious in climate mitigation than others, climate policy can be abruptly disrupted, influencing voting behaviour. In this study, we analyze the political feasibility of a set of green policies in case some parties are strongly unwilling to protect the climate. We show that this simple additional social layer of complexity largely affects the outcome of the abatement measures.
In particular, we conclude that a (high) pure carbon tax is particularly vulnerable to abrupt interruptions and its economic side effects discourage votes for green parties. Nevertheless, a strategically selected combination of policies can reduce political uncertainty, resulting in a more feasible and effective mitigation measure.
\end{abstract}


\section{Introduction}
Climate mitigation pathways are currently explored in so-called Integrated Assessment Models (IAMs) \cite{Dowlatabadi1995, Weyant2017}. IAMs are elaborate tools but generally have two weaknesses: they have difficulties modelling abrupt changes, and they focus only on specific subsystems, namely the economy (especially the energy and land-use sector) and the climate while ignoring social and political processes \cite{Keppo2021, VanSoest2019}. These models evaluate climate policies from an economic perspective, typically focusing on cost-effectiveness \cite{Krey2014, Fisher2020}. \\
\indent However, the economy is not an isolated system. Rather, it is intertwined with the political system, in which policymakers impose economic policies, but are (in democracies) dependent on public opinion, which in turn can be influenced by economic performance \cite{Peng2021, Meng2019}. Changing governments can lead to disruptions in climate policy if some parties are much more ambitious in climate protection than others. On the other hand, climate policy will also influence voting behaviour. In light of these considerations, previous research has emphasized that the primary source of uncertainty in curbing global temperature rise is less about scientific or technological issues and more about political decisions that postpone mitigation efforts \cite{Rogelj2013}. Therefore, a truly integrated assessment of climate policy must take into account socio-political processes, including a political system as well as human and collective behaviour \cite{Beck2016}. Decision-makers with climate ambitions need to find policies that are feasible and robust against political instability and interruptions. A recent example of a politics-induced failure is the fuel tax proposed by the French government in 2018 which caused a large series of protests around the country and subsequently a big step backwards by the policymakers. \\
\indent In this study, we assess the effect of party politics on the green transition by implementing a 
two-party election model in an agent-based IAM to analyze the effectiveness of a set of green policies in case of one party is unwilling to protect the climate. \\
\indent Three election scenarios are compared: No elections (``green'' party always rules), Coinflip (the ruling party is selected with 50\% chance every 4 years) and Coupled (chance of re-election depends on economic performance and global warming). We model public discontent with (economically costly) climate policies by assuming that voters punish parties during whose government unemployment rises. This is based on empirical evidence from various countries that suggests that experiencing job loss often precipitates a shift in the political spectrum \cite{Sipma2023, Wright2012, Wiertz2021}. \\
\begin{figure}[h]%
    \centering
    \includegraphics[width=\linewidth]{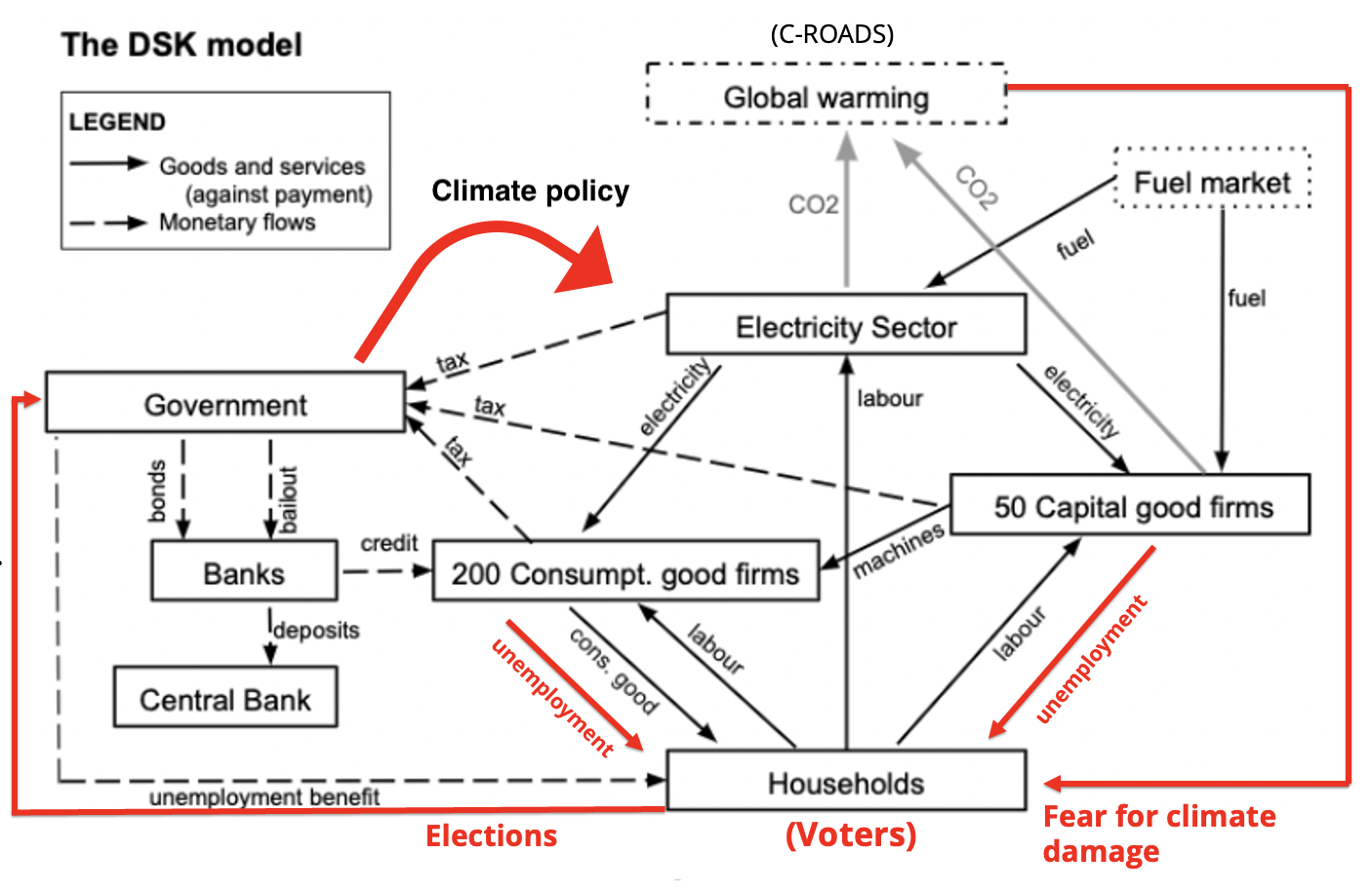}
    \caption{In black the processes of the previous version of the model \cite{Wieners2022, Lamperti2019}. In red, are the newly introduced socio-political feedbacks. Climate policies, which were present also in the previous model version, are now implemented by parties whose probability of being elected depends on the state of the economy and climate.} 
    \label{sketch}%
\end{figure}
\indent As the Integrated Assessment Model, we use the Dystopian Schumpeter-Keynes (DSK) model, the first agent-based IAM \cite{Lamperti2019}. The DSK model simulates the industrial sector of one country. 
Its agent-based nature allows for a detailed analysis of the macroeconomic transition costs, which are often overlooked by models based on General Equilibrium (GE) assumptions.
In DSK, complete decarbonisation requires a transition to green energy in the electricity sector and full electrification in the heavy industry (``capital goods'') sector. 
Here, we test several policy mixes (see Table \ref{tablepolicies} and Methods), combining carbon tax, subsidies and regulations, for their political feasibility under elections. To this end, we define the difference in temperature outcomes between the No-election and the Coupled case  as \emph{political uncertainty} \cite{Rogelj2013, Vaughan2009, Bosetti2009, Sanderson2020, Beck2016} and consider policies which minimise political uncertainty to be \emph{politically feasible}. \\
\indent Throughout this study, our aim is to explore the overall system behaviour resulting from the complex interactions among economics, climate and society, rather than to provide exact quantitative predictions.

\section*{Main}
\label{sec:headings}
\subsection*{Pure carbon tax}
\begin{figure}%
    \centering
    \includegraphics[width=\linewidth]{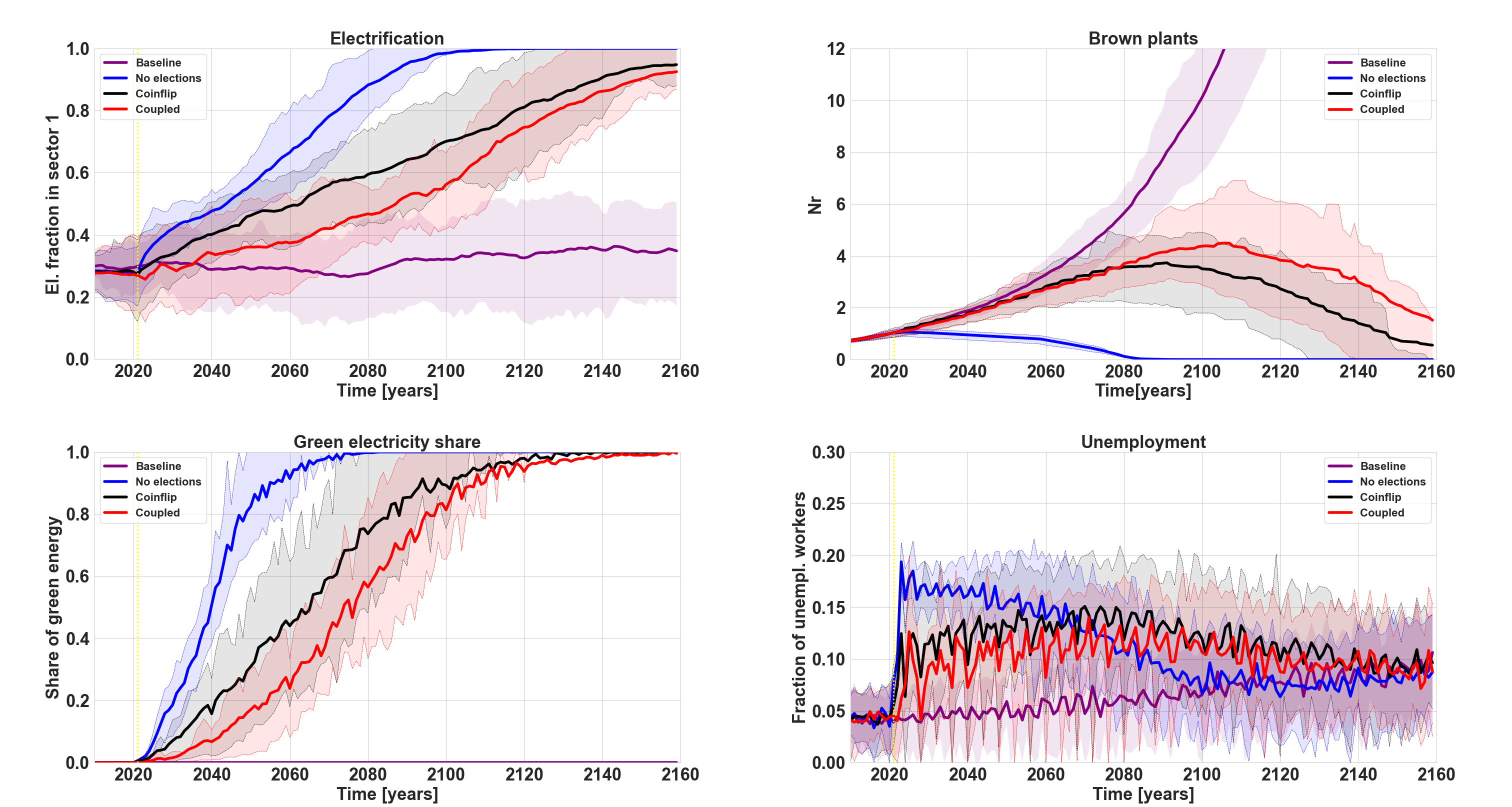}
    \caption{Timeseries of the electrification fraction in the capital good sector, the number of brown plants, the fraction of unemployed workers and the share of green energy for the three election cases and policy T2 (See Table \ref{tablepolicies}). The shaded region represents the range from the 20th to the 80th percentile, providing a view of the variability in outcomes. The thick lines describe the means across 50 Monte Carlo runs.  The simulation starts in 2000. The number of plants normalized by the 2020 mean value.}
    \label{processes1}%
    \qquad
    \includegraphics[width=\linewidth]{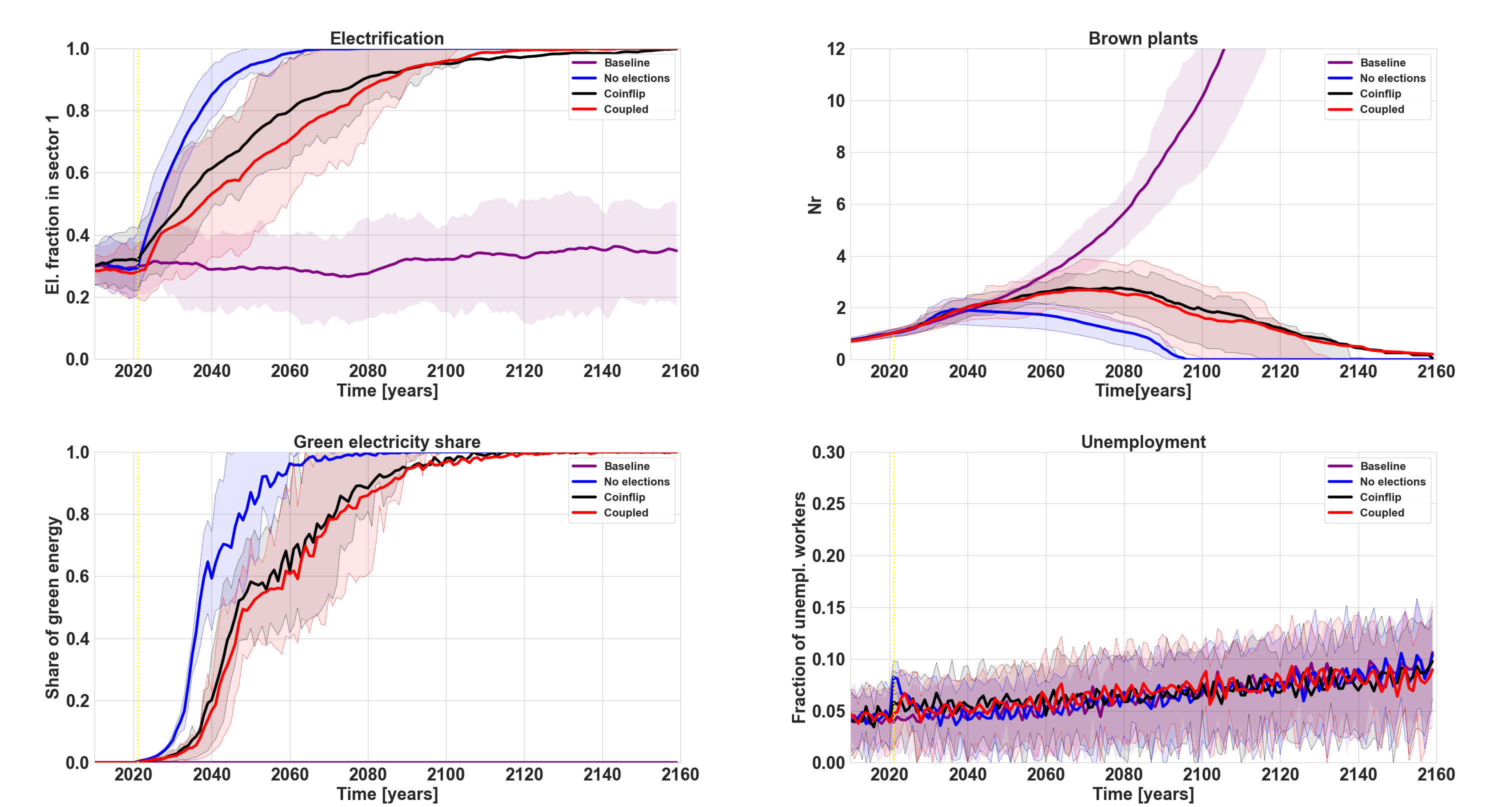}
    \caption{Timeseries of the electrification fraction in the capital good sector, the number of brown plants, the fraction of unemployed workers and the share of green energy for the three election cases and policy T2eB (See Table \ref{tablepolicies}). The shaded region represents the range from the 20th to the 80th percentile, providing a view of the variability in outcomes. The thick lines describe the means across 50 Monte Carlo runs.  The simulation starts in 2000. The number of plants normalized by the 2020 mean value.}
    \label{processes2}%
\end{figure}

\begin{figure}[h]%
    \centering
    \includegraphics[width=\linewidth]{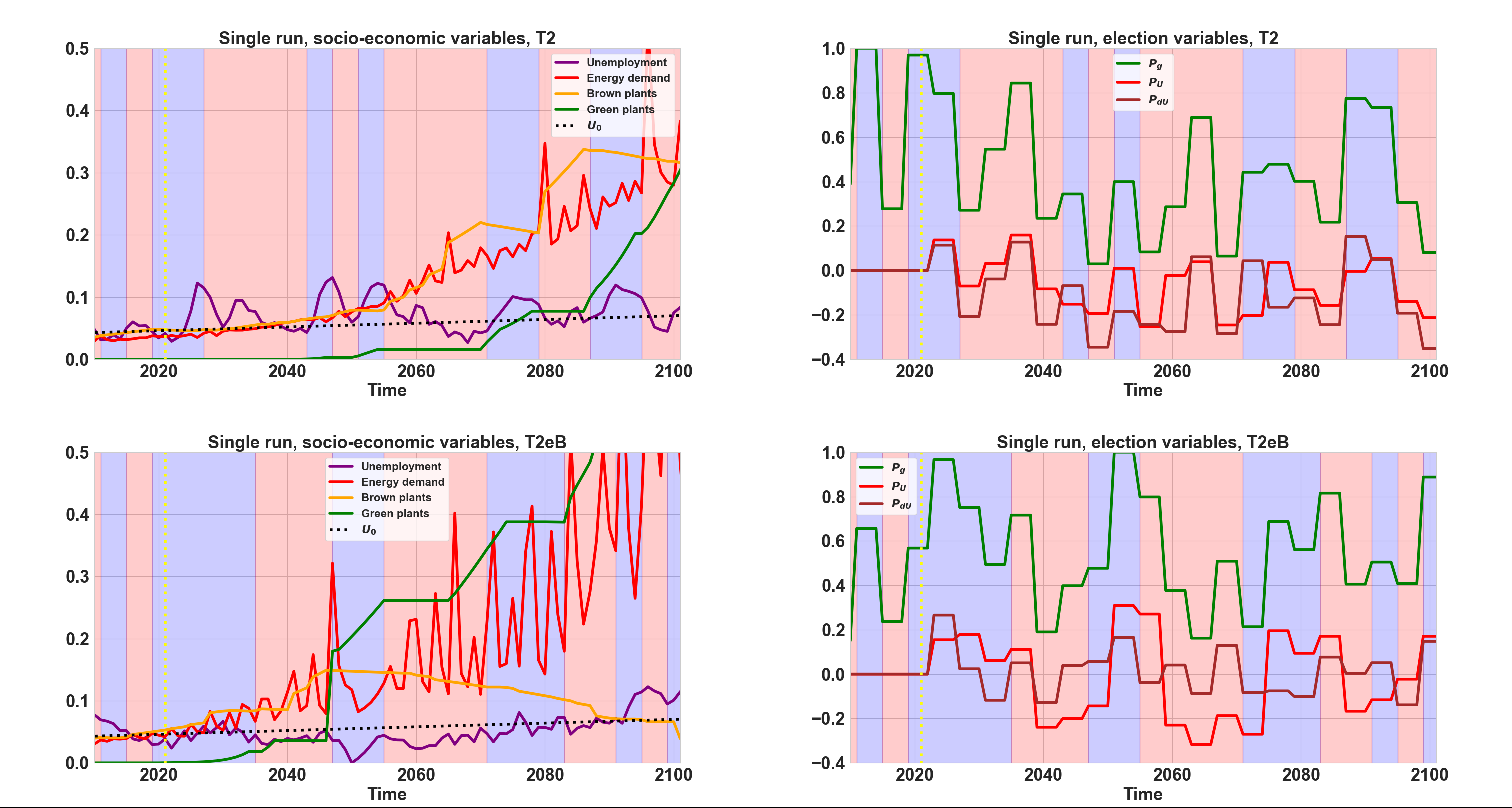}
    \caption{Overview of the processes arising in the coupled case when a carbon tax T2 or a combination of sectoral tax and brown investment ban (T2eB) are implemented by the green party. The vertical yellow line marks the start of the implementation of climate policies. The left plots show the evolution of meaningful macro-variables in a single run. The number of plants normalized by the 2120 baseline mean value. On the right plots, the evolution of the terms involved in the election model (See Eq. \ref{elections}) from the same simulation. $P_{g}$ is the probability for the Green Party of being elected. Blue (red) shaded means Green (Brown) party in power. For example, with T2, the Green Party is in power during the 2022-2025 mandate (right plot), introducing a carbon tax causing unemployment to grow until 2025 (left plot). This leads to a drop in `$P_{U}$ and $P_{dU}$ (See Eq. \ref{elections})
    for the mandate 2026-2029 (right plot), lowering the chance of Green being elected to 0.3. Indeed, the brown party is elected in 2026. Unemployment is a 4-year moving average.}
    \label{singlerun}%
\end{figure}

In the absence of elections, a moderate carbon tax (Tc, see Table \ref{tablepolicies}) ensures a green transition in the electricity sector. However, to keep warming well below 2°C, a much higher tax (T2) is needed \cite{Wieners2022}, because of inertia in the electricity sector \cite{Grubb2021} and because the tax increases both fuel and electricity price; thereby being slow in enforcing electrification in the capital-good firms. The T2 carbon tax leads to a strong rise in fuel and electricity prices and higher production costs for the capital goods sector. Machine sales and production hence drop, leading to unemployment. Higher production costs and lower employment (and hence, lower consumption) also lead to slight reductions in consumption good production, exacerbating unemployment. This causes a decade-long unemployment rise (17.6\% for 2024-2028; baseline: 4.4\%; Fig. \ref{processes1}). \\
\indent In the Coinflip model, both green technologies in the electricity sector and electrification are significantly delayed (Fig. \ref{processes1}), leading to 3 degrees warming, compared to 1.8 in the No-election case (Fig. \ref{overview}).  
This is because in years when the Brown party is in power, no carbon tax is in place to prevent the construction of brown plants (Fig. \ref{singlerun}). In fact, in such years the economy recovers from the stagnation induced by the tax, boosting the energy demand and hence (brown) plant construction.\\
\indent In the Coupled case, mitigation results are further worsened because high peaks in unemployment during the Green party's mandate reduce the probability of the Green party being elected (Fig. \ref{singlerun}). These unemployment spikes stem from the implementation of a tax policy during the Green Party's mandates, which lower $P_{U}$ and $P_{dU}$ (See Methods, Eq. \ref{elections}), thus the probability of being elected $P_{g}$ (See Fig. \ref{singlerun}). On average, the Green Party has a <50\% chance of being in power in 2020-2100 (See Fig. \ref{processes1}, \ref{singlerun}). This leads to a poor performance in limiting global warming, with 3.5 degrees warming by 2100 (See Fig. \ref{overview}). 

\subsection*{Pure carbon tax with subsidy}

By redistributing the tax revenue, the negative impact of the carbon tax on the unemployment rate can be significantly reduced. A possible approach could be redistributing the money as a wage subsidy to firms (T2f). This measure substantially lowers the unemployment peak to approach baseline levels, with a slower green transition which results in similar maximum levels of global warming (around 3 degrees) and political uncertainty for both the election cases (See Fig. \ref{overview}). This removes the punishment faced by the Green Party and narrows the gap in temperature projections between the Coinflip and Coupled scenarios (See Fig. \ref{overview}). Another possibility is subsidizing the construction of green plants (T2C). This measure does not directly reduce unemployment in the short term (20\% peak, See Fig. \ref{overview}) but promotes a faster green transition. This leads to reduced maximum temperatures for all election cases, improving its effectiveness but not its political feasibility.

\subsection*{No-tax policies}

\begin{figure}[h]%
    \centering
    \includegraphics[width=\linewidth]{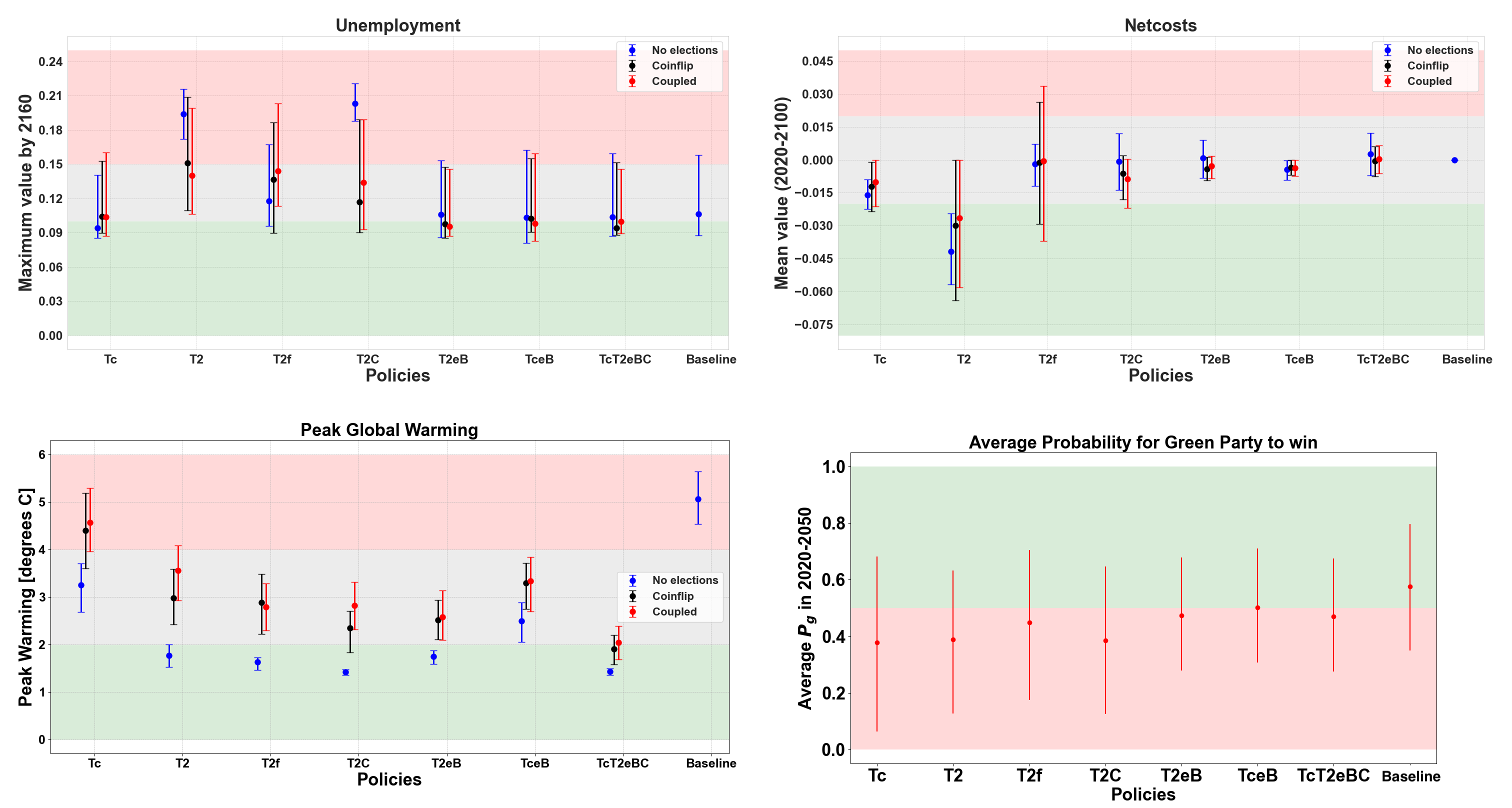}
    \caption{Overview of the results for the policies discussed (See Table \ref{tablepolicies}). 
    All values are averaged across 50 Monte Carlo runs and error bars reproduce the spread between the 20th and 80th percentile of the ensemble. Global peak warming in degrees Celsius. Netcosts in portion of GDP, averaged across the time interval (2020-2100). The average $P_{g}$ is shown only for the coupled case.}
    \label{overview}%
\end{figure}

Another possible strategy is the implementation of investment restrictions or incentives to trigger the construction of new green plants together with regulations to enforce the electrification of the capital good sector. Within this model, electrification can be enforced by a sector-specific carbon tax on the "hard-to-abate" capital good industry.
We consider two different intensities, T2e and Tce. For the high tax, T2e, 80\% electrification is reached after a decade, while for the weaker tax, Tce, it takes a decade. 
In the electricity sector, we consider a ``brown ban'' which prohibits further investments in brown plants (B). In the absence of elections, the brown ban becomes active 16 years after its announcement in 2020. Within this grace period, the electricity firm can ``prepare'' by investing in green RnD and already building up a stock of green plants \cite{Wieners2022}. When the brown party takes power, it revokes the announcement of the ban, and the electricity firm stops preparations.
Therefore, the actual enforcement of the ban is postponed by the number of years that the brown party holds power during this 16-year grace period. After the enforcement starts, the ban is only enacted when the Green Party rules, while under the Brown Party, the construction of brown plants remains possible. \\
\indent With the sectoral tax only addressing electrification and the brown ban only green electricity production, a combination is needed to decarbonise. 
Under No-elections, the combination T2eB achieves similar mitigation to T2 (See Fig. \ref{overview}), but when elections are present, it outperforms T2. Unemployment rates are likewise lower under T2eB. \\
\indent The combination T2eB achieves similar mitigation to T2 alone, while lowering the unemployment rate peaks (See Fig. \ref{processes2}). In the electricity sector, two factors contribute. First, during the years when the electricity form ``prepares'' for the ban, large green investments are made, which directly increase green electricity and also help to drive down the price of green plants. Second, T2eB does not cause strong economic stagnation when the Green party rules, and therefore does not strongly reduce the incentive to build new plants precisely while policies are in place to ensure green investments (See Fig. \ref{singlerun}). 
By combining T2eB with a modest carbon tax (Tc) that is used to subsidise further green plant construction (policy mix TcT2eBC), we obtain the fastest green transition with the lowest political uncertainty and comparatively low unemployment (See \ref{overview}).\\
\indent The result by Wieners et al. \cite{Wieners2022} that a strategically selected combination of policies leads to the best outcomes thus also holds in the presence of elections. In particular, a well-chosen policy mix can strongly reduce political uncertainty w.r.t. a pure carbon tax.

\subsection*{Discussion and conclusion}

We illustrated the complex dynamics which arise by encapsulating political dynamics into IAMs. More specifically, we describe an idealized two-party scenario in which the Green party introduces ambitious climate policy while the Brown party blocks them all. In our model, either party can instantaneously fully revert the other's policy, which is of course a simplification \cite{Basseches2022}.\\ In our framework, the success of incumbents hinges on their handling of unemployment and climate change mitigation. However, real-life elections are influenced by a wider range of factors, including candidate likability, campaign strategies, regional dynamics, and exogenous events like wars, terrorist attacks, or pandemics. In addition, our current model version assumes bulk probabilities for re-election rather than modelling opinion dynamics explicitly. Future work could integrate a more advanced agent-based framework that accounts for the diverse (voting) behaviours of heterogeneous households  \cite{DiBenedetto2023, Lackner2024}. Another possible limitation is the fact that policy costs borne by the state - such as refinancing the electricity firm in case it goes bankrupt under a brown ban - have little effect on the wider economy, as the state is not cash-constrained and no adverse consequences follow from increased spending on climate policy. However, in real life, a modest carbon tax could be levied to fund climate policy spending. \\ 
\indent In conclusion, we have presented a modelling framework for incorporating political dynamics into integrated assessment models.
Our results suggest that a high pure carbon tax, which could seem effective in reducing emissions, is particularly susceptible to abrupt interruptions caused by democratic alternation.
These "Brown" periods coincide with increased energy demand which causes boom phases of brown plants construction delaying the green transition.
On top of this, when coupling elections to parties' performances, results get even worse with the Green Party punished by its strongly unemployment-affecting policies, demonstrating that a carbon tax does not represent a feasible policy. This high political uncertainty poses a significant challenge to the implementation of high carbon taxes in democracies and necessarily asks for the implementation of other additional policy tools. We show that a viable solution to this final complication is the redistribution of the tax revenue. Another promising approach is combining strategies aimed at both pushing for the electrification of the capital good sector and subsidising the construction of new renewable green power plants. In this scenario, the electricity sector is forced to build green plants and the negative impact of the elections is mitigated. \\
\indent These findings underscore that politics plays a crucial role in policy assessment. A more nuanced and politically aware policy design for climate change mitigation strategies is necessary, with policies that must be robust to disruptions and publicly acceptable \cite{Hammerle2021, Kallbekken2013}: their objectives, mechanisms, and impacts, should be readily available and comprehensible to voters. Only in this way, the public can engage with, understand, and support them, contributing to their long-term endurance and effectiveness with their votes. We aim to spark the interest of fellow researchers, encouraging them to build on our findings in more intricate and expansive models. This effort marks a step toward enriching the understanding of complex interactions within climate policy assessments.

\section*{Methods}

\subsection*{The original DSK model} 
We introduce a 2-party election model in the Dystopian-Schumpter-Keynes model \cite{Lamperti2018, Lamperti2019} where one party with strong climate policy ambitions competes with a party with low ambition (see Fig. \ref{sketch}). 
The DSK model an agent-based IAM which describes the economy of a country by modelling a set of heterogeneous, interacting firms and a climate module.
50 Capital-good firms produce machines, using labour and energy. The energy can be fuel, which is bought on an exogenous fuel market, or electricity, which is provided by the electricity sector.
200 consumption-good firms buy machines and use them, as well as labour and electricity (but not fuel), to produce a homogenous consumption good that is sold to households.
Households are modelled as an aggregate labour and consumer force. They work, buy consumption goods from the market and, if applicable, receive unemployment benefits from the government.  Electricity is produced by a profit-seeking monopolist using ``brown'' (fuel-burning) and ``green'' (renewable) electricity plants. Plants are scrapped after 60 years. New ones are built such that the electricity demand by capital-good and consumption-good firms is always fulfilled. Green plants initially have high fixed costs, brown ones have low fixed costs but operational costs from fuel use. Without climate policy, green plants never become cost-competitve. Research and Development (RnD) can reduce fixed costs for both types of plants, and improve fuel efficiency in brown plants. Innovation potential is greater for green plants, as brown ones are already a mature technology, so green plants can eventually become cost-competitive. The RnD budget is split between brown and green plants according to their fraction in energy production; in particular, if green plants are not used, they receive no RnD funding.
Decarbonisation requires 1) full green energy in the electricity sector and 2) electrification in the capital good sector.\\
\indent Climate dynamics follow the C-ROADS model \cite{Sterman2012}. 
When CO2 is emitted, part of it remains in the atmosphere and leads to global warming. The equilibrium climate sensitivity (ECS) is 3 degrees (Kelvin) per doubling CO2.  
We do not take climate damage into account, as DSK lacks many of the most vulnerable sectors (e.g., agriculture), and our focus is on how to avoid dangerous climate change, rather than studying its impact. In the no-policy baseline, GDP growth is 3\% (2020-2120) and global warming is 4.8 degrees in 2100 (see Fig. \ref{overview}), in line with the SSP5 scenario \cite{Kriegler2017}. The economic model is spun up for 60 years before the actual simulation. The climate model is activated after spin-up and initialised to observed 2020 conditions. Climate policy starts in 2021.  The model has been shown to reproduce key emergent properties of the economy (“stylized facts”) \cite{Dosi2010, Dosi2013, Lamperti2018} and allows to study of the dynamic impacts of a rich set of climate policies \cite{Wieners2022}. 

\subsection*{Coupling DSK to a simple election model}
In our simple election model, the ``green'' party implements an ambitious climate policy, while the ``brown'' party implements no climate policy at all.  The electorate votes based on the parties' handling of climate policy and the economy. \\
To distinguish between the effects of (random) interruptions of policies and actual election dynamics, we compare two election models. In the ``coinflip'' model, the ruling party is elected randomly, with equal probabilities, every four years. In the ``coupled'' model, the probability $P_g$ of the Green Party to be elected for a mandate $M$ is given by

\begin{equation}
  P_{g}(M) = 0.5 + P_{T} + P_{U} + P_{dU} + P_{B},
\label{elections}
\end{equation}

where 

\begin{equation}
\begin{split}
&P_{T} =  \alpha T, 
\\
&P_{U} = - \sigma \beta[(U(M-1) - U_0)], 
\\
&P_{dU} = - \sigma \gamma[U(M-1) - U(M-2)],
\\
&P_{B} =- \sigma \epsilon N.
\end{split}
\end{equation}

The term $P_T$ reproduces an inclination to vote green when global warming ($T$) is felt, with $\sigma=+1$ if the Green Party is incumbent and -1 else.
$\alpha>0$ is chosen such that a 3-degree warming results in an advantage of $12\%$ for the Green party.\\
\indent $P_{U}$ and $P_{dU}$ reproduce the influence of unemployment on the probability of electoral success.
$P_U$ depends on the difference between the average unemployment rate from the previous election cycle $M-1$ and the average unemployment in a no-policy baseline scenario $U_{0}$. Similarly, $P_{dU}$ reflects the change in average unemployment rates across the two most recent electoral terms, Voters thus punish the incumbent party if recently unemployment was high ($P_{U}$) or got worse ($P_{dU}$). The parameter $\beta>0$ is calibrated on historical data for presidential odds \cite{SportsOddsHistory2024} and the unemployment rate change in the U.S. \cite{WikipediaUserTalk2024} (see SM, Fig. 1-2). For simplicity, we assume $\beta=\gamma$. \\
\indent Several studies documented that unemployment adversely impacts both the personal well-being and professional lives of individuals. This often influences political leanings and voting patterns over time. Observations from different countries indeed show that job loss often triggers an ideological response towards the party historically considered more suitable for solving the issue \cite{Sipma2023, Wright2012, Wiertz2021}). Similarly, we assume that unemployment triggers a response favouring the party deemed most likely to mitigate these issues, which, in our simulation, is the Brown one. \\
\indent $P_{B}$ (``the boredom term'') describes voters gradually getting fed up with the ruling party and is proportional to the number of consecutive mandates the incumbent party has ruled before the elections. Empirical observations suggest that more consecutive mandates mean a lower probability of being elected, preventing the system from reaching a state where one party has a persistent overwhelming majority \cite{Fair2009}.
However, the exact degree of influence ($\epsilon$) of this pattern on electoral outcomes remains partially uncertain, as these qualitative insights do not directly translate into specific electoral odds (see SM, Fig. 3). \\
\indent The probability of being elected for the Brown Party is given by $P_{b}(M) = 1 - P_{g}(M)$. 
Throughout the study, we refer to this final modelling setup as the "Coupled case". The calibration, along with sensitivity analysis, is available in the Supplementary Material.

\begin{table}[h]
\centering
\begin{center}
\begin{tabular}{||p{2cm}| p{12cm}||} 
 \hline
 \textbf{Policy} & \textbf{Explanation} \\ [0.5ex] 
 \hline 
  Baseline & Baseline experiment without any climate policy.\\ 
 \hline 
 Tc & Critical tax: constant carbon tax, just sufficient to induce a green transition in all Monte Carlo members. It starts in 2022 and the revenue is added to the government budget. \\  
 \hline
 T2 & Tax high enough to keep global warming below 2 degrees until 2160 in all Monte Carlo members without any additional policy. \\
  \hline
 T2f & As T2, but tax revenue is spent on subsidising firms to pay wages. \\
  \hline
    Tce & Sector-dependent tax: As Tc, but the tax applies only to the capital good sector, to promote electrification. \\
 \hline
 T2e & Sector-dependent tax: As T2, but the tax applies only to the capital good sector, to promote electrification. \\
 \hline
 B & Brown investment ban. Brown plants cannot be built 16 years later than the announcement. \\
 \hline
 C & The state invests the money obtained by the carbon tax in building green plants, even if not strictly needed to fulfil demands. \\
 \hline
 TcT2eBC & Tc only applied to the electricity sector combined with T2e, B and C.\\
 \hline
\end{tabular}
\caption{Climate policies. Combinations are named by directly appending the names of the individual policies it comprise. }
\label{tablepolicies}
\end{center}
\end{table}

\section*{Acknowledgements}
This publication is part of the project ‘Interacting climate tipping elements: When does tipping cause tipping?’ (with project number VI.C.202.081 of the NWO Talent programme) financed by the Dutch Research Council (NWO).
The authors are thankful to Teresa Lackner, Simon L. L. Michel,  Luca Fierro and Patrick Mellacher for helpful comments and discussions.

\bibliographystyle{unsrtnat}
\bibliography{references}



\section*{Author contributions statement}

C. E. W. , A. D. and A. S. H. conceived the experiment;  A. D. conducted the experiment;  all authors reviewed the manuscript. 





\end{document}


\maketitle

\section{The election model}

\subsection{Calibration}

\begin{figure}[h!]%
    \centering
    \includegraphics[width=\linewidth]{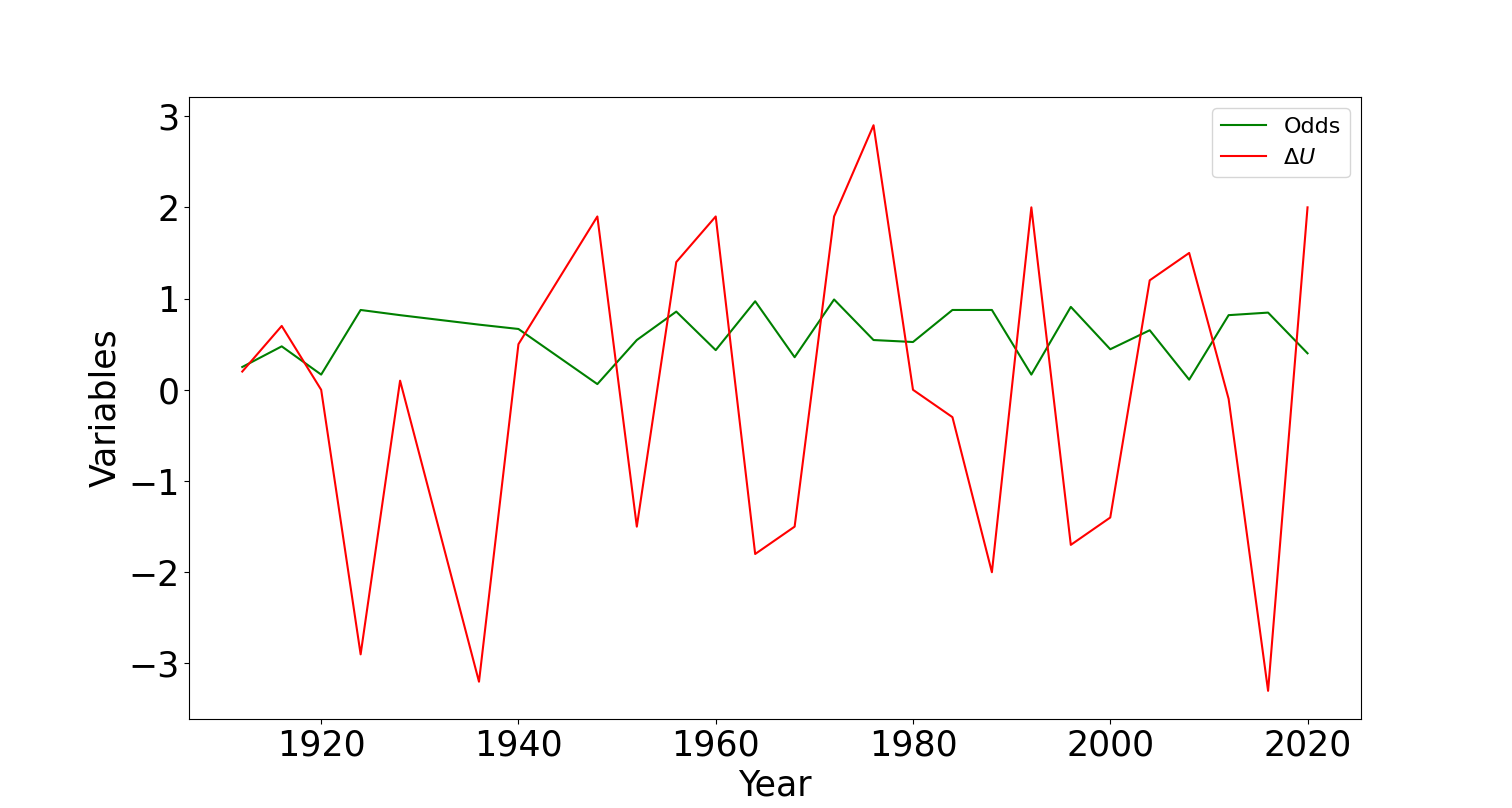}
    \caption{Historical trends in U.S. elections: Odds of the incumbent party's candidate winning (green) vs. changes in the unemployment rate (red), where $\Delta U$ represents the difference in unemployment rates between Election Day and Inauguration Day of its mandate.}
    \label{sketch}%
\end{figure}

\newpage
\begin{figure}[h!]%
    \centering
    \includegraphics[width=\linewidth]{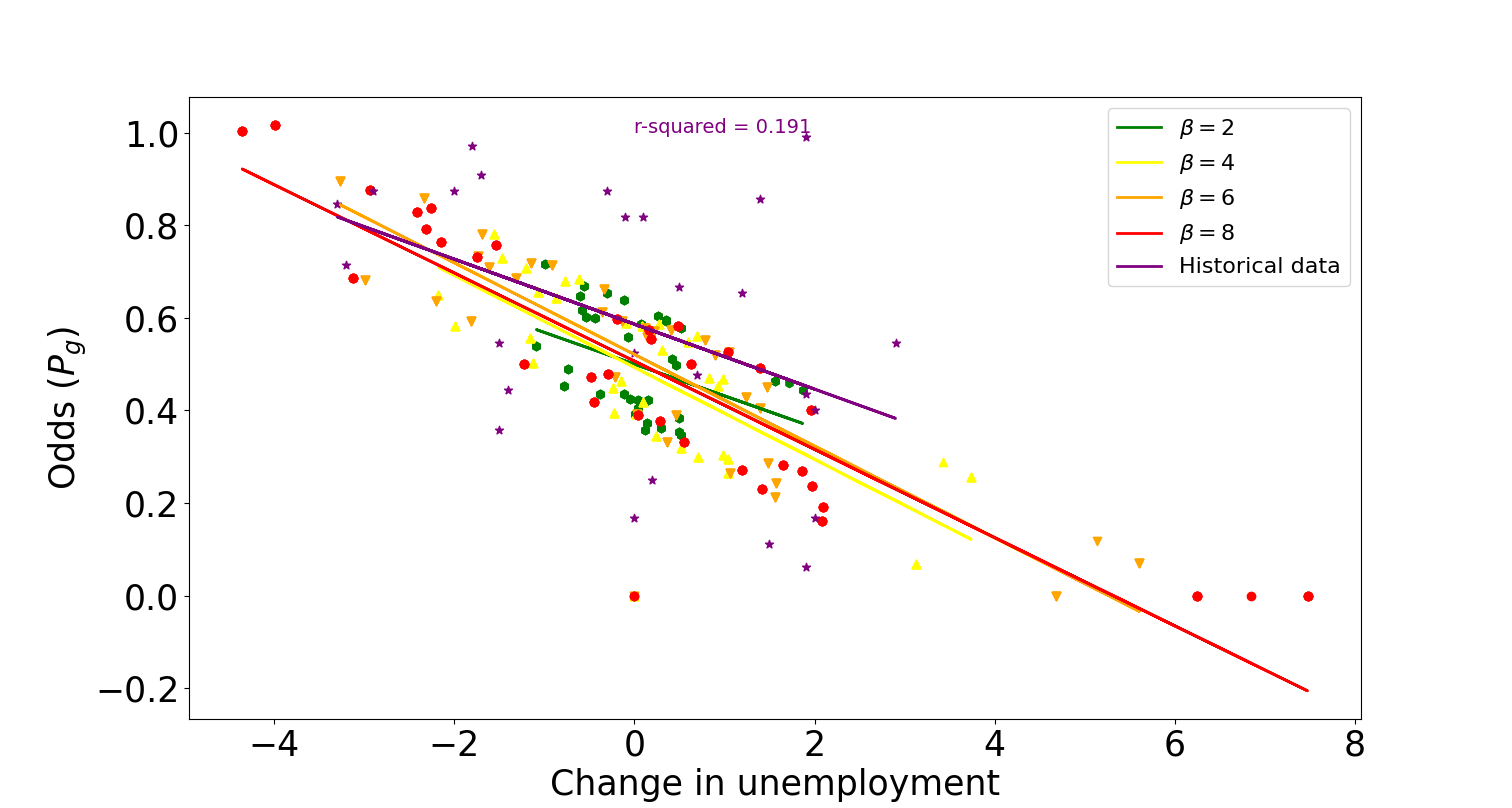}
    \caption{Linear regression analysis of the incumbent party's odds and unemployment rate changes, comparing historical data with model simulations. Elections in 1944 and 1932 have been excluded due to extraordinary circumstances. The x-axis for model simulations corresponds to $P_{U}$. Model parameters: $\alpha=4$, $\epsilon=0.04$ under a 3-degree warming scenario. For our modelling setup, we choose $\beta=6.$}
    \label{sketch}%
\end{figure}

\newpage
\begin{figure}[h!]%
    \centering
    \includegraphics[width=\linewidth]{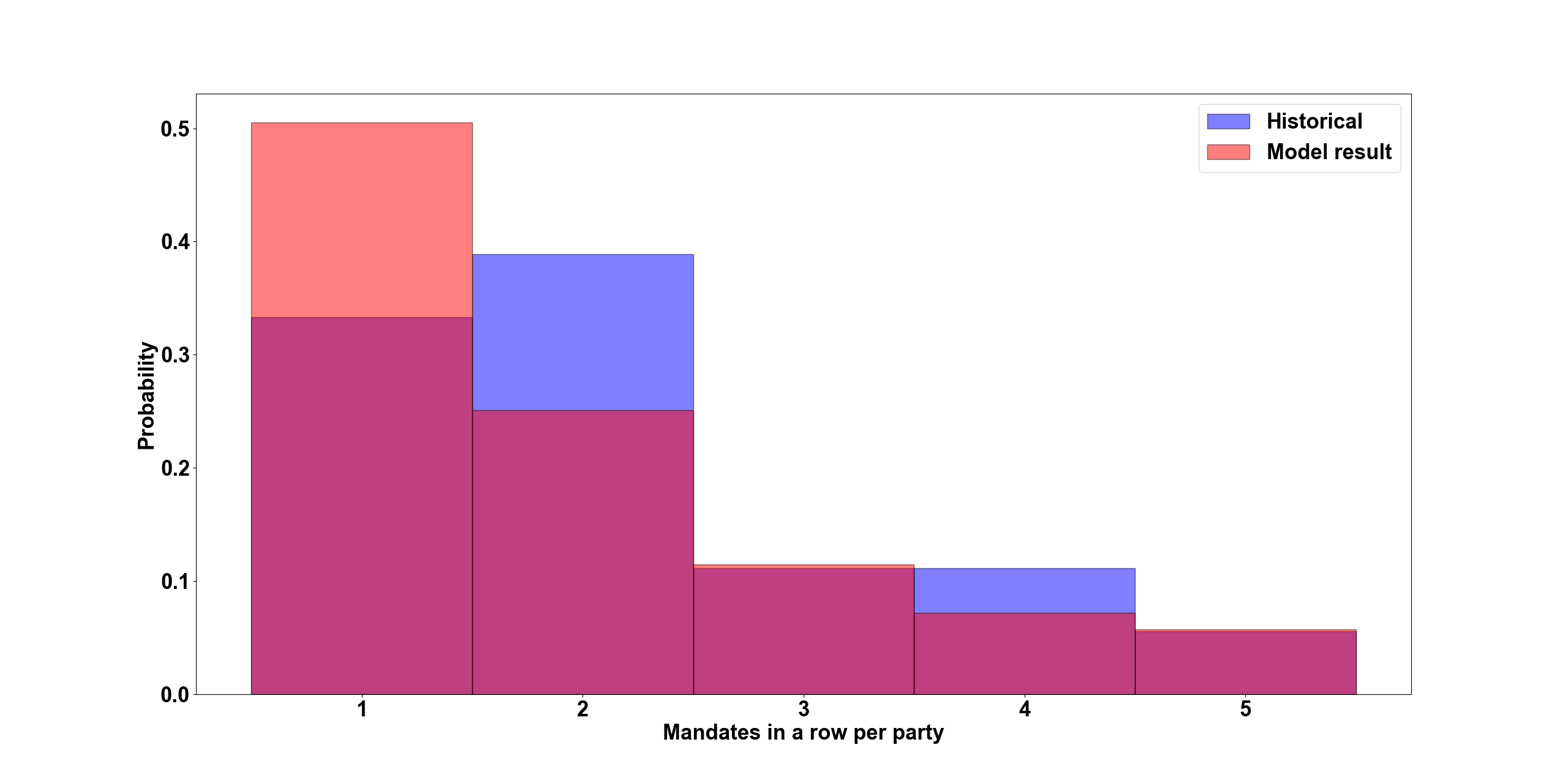}
    \caption{Histogram showing the distribution of consecutive terms per party in U.S. presidential elections in historical data and model results. Simulations are obtained in a 3-degree warming scenario. To fit an average of approximately two terms like historical data, $\epsilon$ has been fixed to 0.04. Model result frequency is evaluated across 50 Monte Carlo runs, only considering the time window 2020-2090.}
    \label{sketch}%
\end{figure}

\newpage
\subsection{Sensitivity analysis}

In this subsection, we discuss how the model is sensitive to the election parameters $\alpha$ and $\beta$. By increasing $\alpha$, the impact of climate change on elections increases, despite the unemployment term being fixed to $6$ (Fig. 4). When changing both, two different regimes can be detected in terms of $P_{g}$ (Fig. 5) and global warming (Fig. 6). High $\alpha$ and low $\beta$ result in a highly "Green" scenario, where the Green party has an advance of 25\% if global warming exceeds 3 degrees and consequent mitigated peak in warming of around 3 degrees by 2160. 

\newpage

\begin{figure}[h]%
    \centering
    \includegraphics[width=\linewidth]{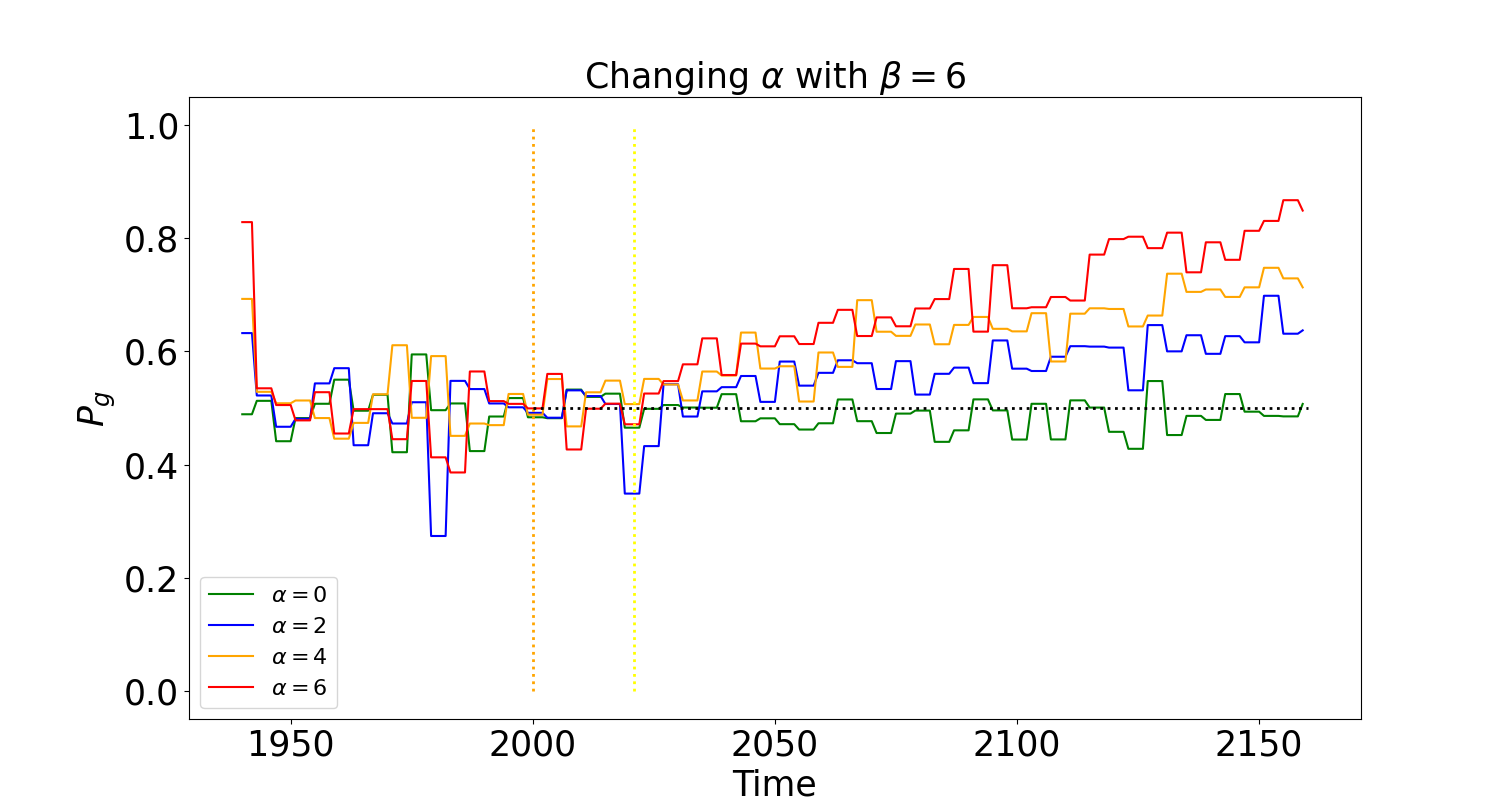}
    \caption{Coupled case. Probability for the Green Party to be elected as a function of time for changing values of the parameter $\alpha$. Each line is a mean value across 50 Monte Carlo simulations. $\beta=\gamma=6$,  $\epsilon=0.04$. The vertical orange line marks the end of the spin-up time. The vertical yellow line marks the start of the implementation of climate policies. }
    \label{sketch}%
\end{figure}
\newpage
\begin{figure}[h]%
    \centering
    \includegraphics[width=\linewidth]{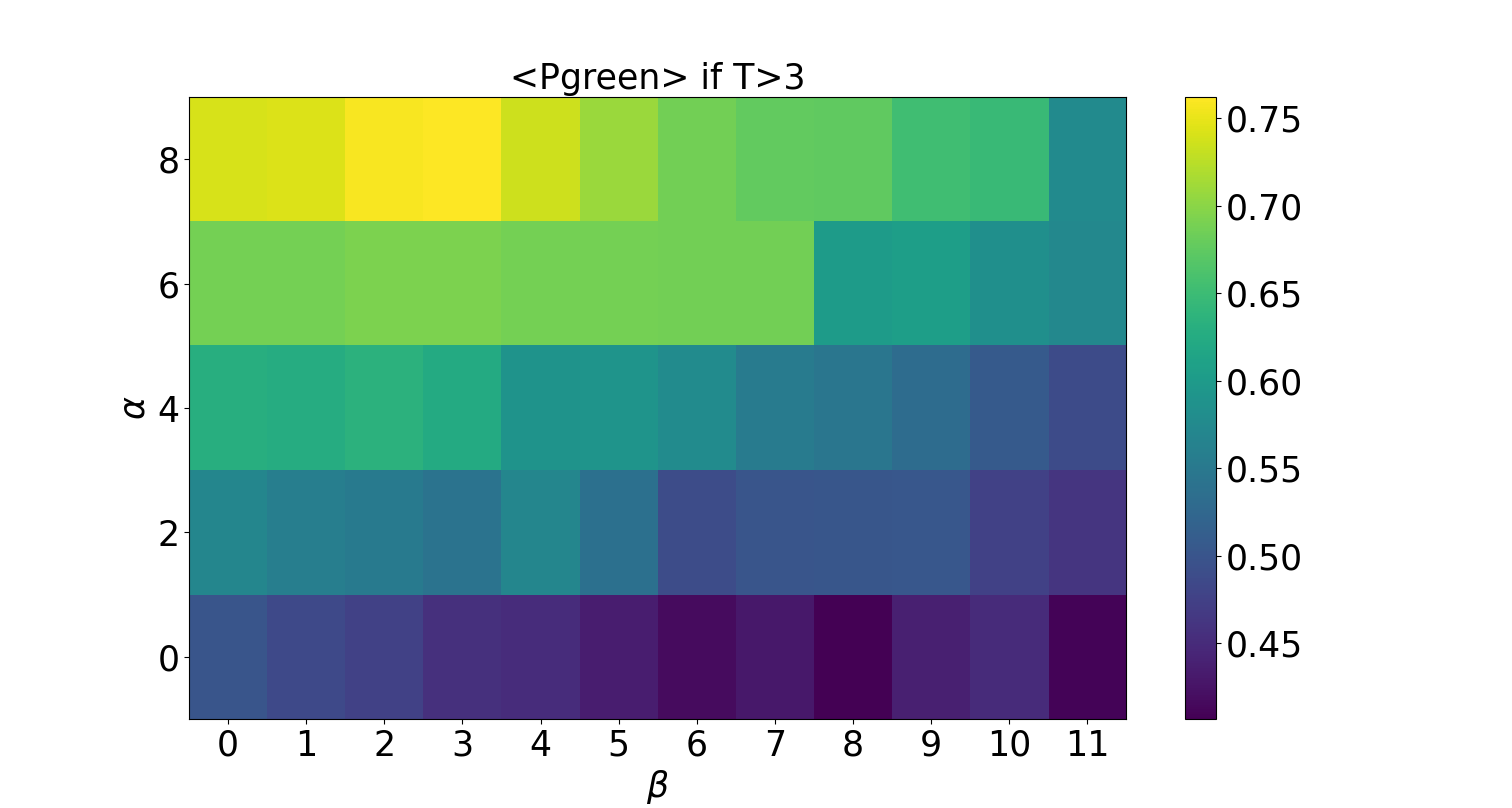}
    \caption{$T2$ in the coupled case. Sensitivity of the time-average $P_{g}$ after the temperature reaches the threshold of 3 degrees for different combinations of $\alpha$ and $\beta$. Mean value across 50 Monte Carlo simulations. For our experiments, we chose $\alpha=4, \beta=\gamma=6$. $\epsilon=0.04$.}
    \label{sketch}%
\end{figure}
\newpage
\begin{figure}[h!]%
    \centering
    \includegraphics[width=\linewidth]{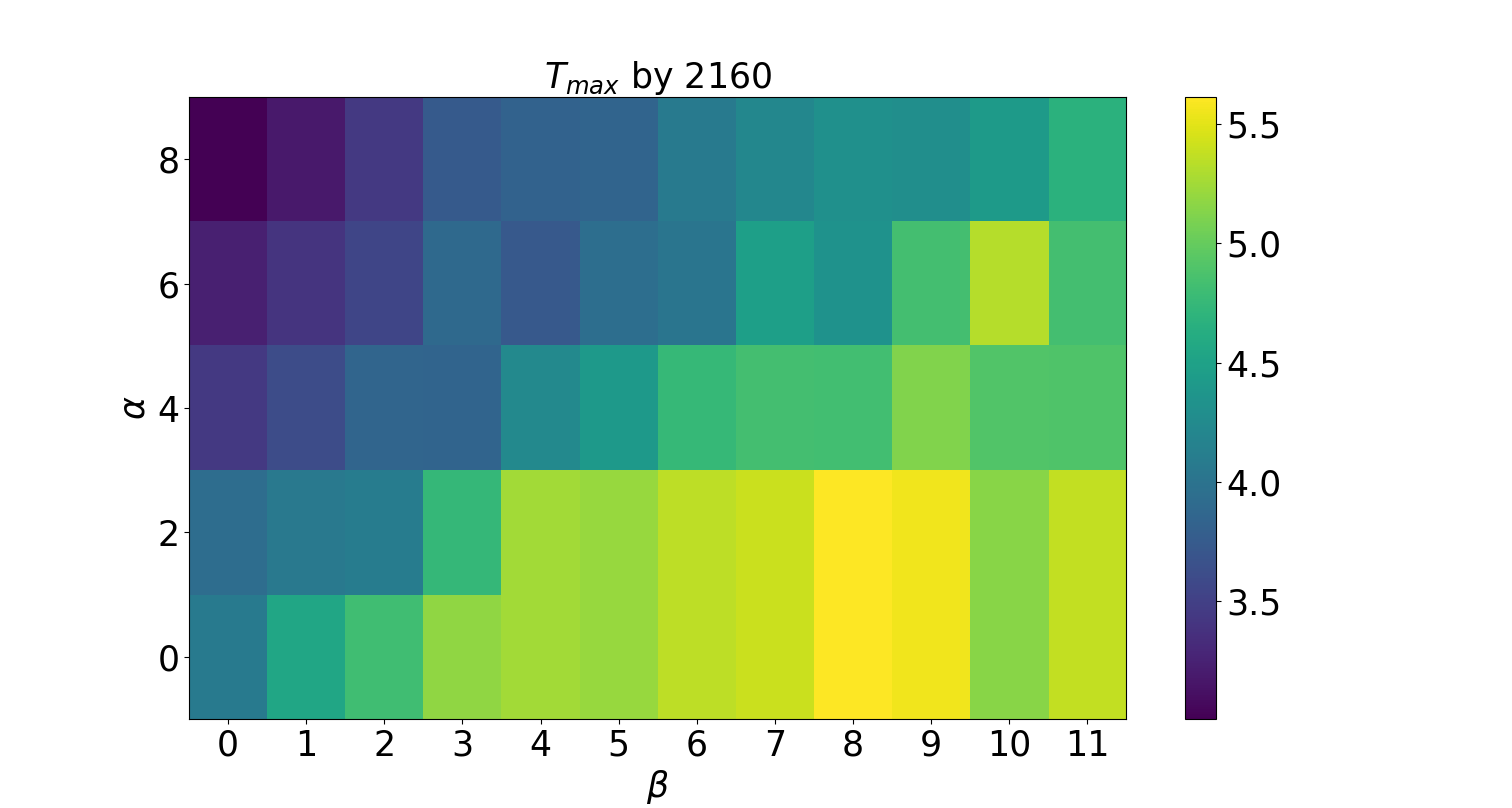}
    \caption{$T2$ in the coupled case. Sensitivity of the temperature response for different combinations of $\alpha$ and $\beta=\gamma$. Mean value across 50 Monte Carlo simulations. For our experiments, we chose $\alpha=4, \beta=\gamma=6$. $\epsilon=0.04$}
    \label{sketch}%
\end{figure}
\newpage



\clearpage
